\documentclass[pre,twocolumn,showpacs,preprintnumbers]{revtex4-1}
\usepackage{amssymb}
\usepackage{amsmath}
\usepackage{graphicx}
\usepackage{color}

\begin{document}

\newcommand{\beq}{\begin{equation}}
\newcommand{\eeq}{\end{equation}}
\newcommand{\beqs} {\begin{displaymath}}
\newcommand{\eeqs} {\end{displaymath}}
\newcommand{\beqa} {\begin{eqnarray}}
\newcommand{\eeqa} {\end{eqnarray}}
\newcommand{\beqas} {\begin{eqnarray*}}
\newcommand{\eeqas} {\end{eqnarray*}}
\newcommand{\reff}[1]{(\ref{#1})}
\newcommand{\derpar}[2]{\frac{\partial #1}{\partial #2}}
\renewcommand{\Vec}[1] {\textbf{#1}}
\newcommand{\Tensor}[1] {\mathbb{#1}}
\newcommand{\Operator}[1] {\mathcal{#1}}
\newcommand{\TensorT} {\mathcal{T\!\!\!\!T}}

\newcommand{\ket}[1]{\vert #1 \rangle}
\newcommand{\bra}[1]{\langle #1\vert}
\newcommand{\braket}[2]{\langle #1\vert #2\rangle}
\newcommand{\mean}[1]{\left\langle #1\right\rangle}
\newcommand{\abs}[1]{\vert #1\vert}
\newcommand{\Li}{{\rm Li}}

\newcommand{\alphak} {\alpha_\Vec{k}}
\newcommand{\betak} {\beta_\Vec{k}}
\newcommand{\gammak} {\gamma_\Vec{k}}
 \newcommand{\Label}[1]{\label{#1}}

\definecolor{magenta}{rgb}{0.7,0,0.7}
\newcommand{\rodrigo}[1]{\textcolor{blue}{[RS: #1]}}
\newcommand{\ricardo}[1]{\textcolor{red}{[RB: #1]}}
\newcommand{\pablo}[1]{\textcolor{green}{[PR: #1]}}

\renewcommand{\theequation}{\thesection.\arabic{equation}}

\title{Dynamical approach to the Casimir effect}

\author{P.~Rodriguez-Lopez, R.~Brito}
\affiliation{Dept. de F\'{\i}sica Aplicada I and GISC,
Universidad Complutense, 28040 Madrid, Spain}

\author{R.~Soto}
\affiliation{Departamento de F\'{\i}sica, FCFM, Universidad de Chile,
Casilla 487-3, Santiago, Chile.}


\begin{abstract}
Casimir forces can appear between intrusions placed in different media driven by several fluctuation mechanisms, either in equilibrium or out of it. Herein, we develop a general formalism to obtain such forces from the dynamical equations of the fluctuating medium, the statistical properties of the driving noise, and the boundary conditions of the intrusions (which simulate the interaction between the intrusions and the medium). As a result, an explicit formula for the Casimir force over the intrusions is derived. This formalism contains the thermal Casimir effect as a particular limit and generalizes the study of the Casimir effect to such systems through their dynamical equations, with no appeal to their Hamiltonian, if any exists. In particular, we study the Casimir force between two infinite parallel plates with Dirichlet or Neumann boundary conditions, immersed in several media with finite correlation lengths (reaction--diffusion system, liquid crystals, and two coupled fields with non-Hermitian evolution equations). The driving Gaussian noises have vanishing or finite spatial or temporal correlation lengths; in the first case, equilibrium is reobtained and finite correlations produce nonequilibrium dynamics. The results obtained show that, generally, nonequilibrium dynamics leads to Casimir forces, whereas Casimir forces are obtained in equilibrium dynamics if the stress tensor is anisotropic.
\end{abstract}
\maketitle

\setcounter{equation}{0}
\section{Introduction}

There are many systems in nature which are subjected to fluctuations, of thermal or quantum origin.
For such systems, under certain physical conditions, Casimir forces, created by the confinement of fluctuations, exist
and have been calculated (see, e.g., \cite{KardarGolestanian}).
The usual way to obtain the Casimir forces uses equilibrium techniques and is therefore valid only for
systems in thermodynamic equilibrium. This means that the fluctuations must satisfy a fluctuation--dissipation
theorem that guarantees the existence of an equilibrium state, as discussed below.
Casimir forces for these systems are calculated in the spirit of the original work of H.\,G.~Casimir for the
electromagnetic case~\cite{Casimir Placas Paralelas}. The method takes as a starting point the Hamiltonian of the
system, from which the partition function $Z=\int \exp(-\beta H)$ is calculated,
either directly or using functional integration~\cite{Kardar-Geometrias-Arbitrarias}. In the calculation of the
partition function one must take into account the boundary conditions, that is, the macroscopic bodies
which are immersed in the system. The partition function of the system will have
different values for
different configurations, e.g., different separations of the objects.
Once the partition function has been obtained, its logarithm provides the free energy $F$.
The final step required to obtain the Casimir force is the calculation of the pressure
as the difference in the free energy when the configurations of the macroscopic bodies change
(for example, changing their position, distance or sizes). For instance, in the usual Casimir case of
forces between two flat parallel plates at separation $L$, the force per unit area is given by
$F_{C}/A=-\partial F/\partial L$.

The second approach also takes as a starting point the Hamiltonian of the system. However, in this approach
the Casimir force is derived not from the free energy but from
the stress tensor $\Tensor{T}$, which is integrated over the surface of the
macroscopic bodies and then averaged over the thermal Boltzmann distribution of the
associated Hamiltonian $\exp(-\beta H)$.
The approach based on the stress tensor has been taken by several authors~\cite{Cardy,Bartolo,Krech,Dean1}.
In fact, both approaches are equivalent and valid for equilibrium systems only.
The reason is that both are based on properties which are only valid in
equilibrium situations. The former uses the thermodynamic relation for the pressure as the derivative of the
free energy with respect to the volume, and the latter uses the Boltzmann distribution
function, which is only valid for systems in equilibrium.

On the other hand, other authors have developed a dynamical approach~\cite{Ajdari,Bartolo,CasimirGranular,Dean2}.
Here the starting point is an evolution equation for the considered field(s), supplemented
with a noise source term, so that the evolution of the field takes the form of a
Langevin equation. Once this equation is solved, the field is inserted into the expression
for the pressure and the average over the noise is taken. As we will see in the next section,
if the noise is of internal origin,
say thermal or quantum, this description reduces to the equilibrium one, because of the fluctuation--dissipation theorem.
However, the noise does not necessarily have to be internal but can have an external origin~\cite{Sagues}, for instance,
a system in a fluctuating temperature gradient~\cite{Gollub}, subjected to
external energy injection such as vibration~\cite{Cattuto} or electrically driven convection~\cite{Brandt}, light incident on
a photosensitive medium~\cite{photosensitive}, or spatially and/or temporally correlated noise, as considered in Ref.~\cite{Bartolo}. Recently, Ref.~\cite{NajafiGolestanian} generalized this method to a nonequilibrium
temperature gradient. In none of these cases can the equilibrium approach
be applied. Also, the internal dynamics cannot satisfy the condition of detailed balance, and therefore the internal noise is not described by the  fluctuation--dissipation relation. In both cases, it is only possible to calculate Casimir forces via the dynamical approach.
In all these nonequilibrium cases a common feature shared with the equilibrium Casimir force is that the origin is the limitation of the fluctuation spectrum at large wavelengths. They are, therefore, conceptually different from other fluctuation-induced phenomena such as ratchets or Brownian motors that act at small length scales.

The plan of the paper is as follows.
We start in Sect.~II by presenting the Langevin equation subjected to a general noise. We stress
the differences between the cases when the Langevin equation derives from an energy
functional or not, and discuss the implications of the fluctuation--dissipation theorem.
Section~III derives the Casimir force from the stress tensor, while Sect.~IV calculates
the actual Casimir force by substituting the solution of the Langevin equation into the stress tensor.

The subsequent sections~V, VI, and VII are devoted to the application of the formalism to different physical systems and different
nonequilibrium conditions, that is, different ways of violating the fluctuation--dissipation theorem.
In particular, Sect.~V studies a reaction--diffusion system with three types of noise: (1)~a noise uncorrelated in
space and time, (2)~a noise exponentially correlated in time, and (3)~a spatially homogeneous noise, only fluctuating in time. Section~VI is devoted to the study of a liquid crystal, with an equilibrium noise, satisfying the fluctuation--dissipation theorem and therefore in an equilibrium situation. We continue with a temporally correlated noise and finish the section with a maximally correlated noise. Finally, to illustrate the power
of the method, we apply it to a two-field system where the evolution equation is non-Hermitian.
Usual approaches, based on equilibrium properties, have no applicability in this case. We finish with some conclusions.
\setcounter{equation}{0}
\section{Equilibrium and nonequilibrium fluctuations}\Label{sec.2}

The most widely used tool to study the dynamics of fluctuations is the Langevin equations and its related Fokker--Planck
equation. There is a wide literature on this subject, in particular using Langevin equations; see, for example, Refs.~\cite{deGroot,Gardiner,Risken,vanKampen,Zwanzig}.

Let us consider a linear stochastic differential equation for the field $\phi({\bf r},t)$,
\begin{equation}\Label{Langevin}
\partial_t\phi=-{\cal M} \phi +\xi({\bf r},t),
\end{equation}
which is a generalization of the Langevin equation to spatially extended systems.
In this equation, ${\cal M}$ is an operator (usually differential)
that can be Hermitian or non-Hermitian.
The operator does not depend on the field $\phi$, so the Langevin
equation (\ref{Langevin}) is linear. To simplify notation, we have assumed Langevin equations without memory, but the generalization to memory kernels is direct.
The term $\xi({\bf r},t)$ is a Gaussian noise that represents the random
or stochastic force acting over the field $\phi$,
and therefore it is the source of fluctuations for $\phi$. It is customary
to assume that the noise is Gaussian, and its averages are
\begin{eqnarray}
\langle \xi({\bf r},t)\rangle&=&0, \Label{noises} \\
\langle \xi({\bf r},t)\xi({\bf r'},t')\rangle&=&{\cal Q} \delta({\bf r}-{\bf r'})\delta(t-t')=h({\bf r}-{\bf r'})\delta(t-t'), \nonumber
\end{eqnarray}
where ${\cal Q}$ is a Hermitian operator that can contain differential and integral terms. Differential terms characterize noise of conserved quantities, and integral terms noises with spatial correlations. The application of this operator to the Dirac delta function produces the spatial correlation distribution $h$.
The
noise here is uncorrelated in time, although temporal correlations will also be considered below.

Equation~(\ref{Langevin}) admits a solution for an initial condition $\phi^0({\bf r})$  as
\begin{equation}\Label{solutionLangevin}
\phi({\bf r},t)=e^{-{\cal M}t} \phi^0({\bf r}) +e^{-{\cal M}t}\int_0^t d\tau \,
e^{{\cal M}\tau} \xi({\bf r},\tau).
\end{equation}
In the limit $t\to\infty$, $\phi({\bf r},t)$ reaches a stationary state if
$e^{-{\cal M}t}\phi^0\to 0$. This implies that the eigenvalues of ${\cal M}$
must have positive real parts.

From the Langevin equation, one can construct a functional
Fokker--Planck equation for the probability distribution $P$
of the field $\phi$. The technique is standard (see, e.g., \cite{Gardiner}), and its solution (which is not normalizable) is a
Gaussian of the form
\begin{equation}\Label{solutionFP}
P[\phi]=\sqrt{\frac{1}{\det {\cal K}}} e^{-\int dr \phi{\cal K} \phi/2},
\end{equation}
where the Hermitian operator ${\cal K}$ is the solution of the equation~\cite{Zwanzig}
\begin{equation}\Label{matrixg}
{\cal M}{\cal K}^{-1}+{\cal K}^{-1}{\cal M}^+= {\cal Q},
\end{equation}
where ${\cal M}^+$ denotes the adjoint of ${\cal M}$.
The probability distribution $P$ depends both on the
matrix ${\cal M}$ and also on the intensity of the fluctuations ${\cal Q}$ via Eq.~(\ref{matrixg}).
However, to the best of our knowledge,
Eq.~(\ref{matrixg}) cannot be solved analytically, so
the operator ${\cal K}$ cannot be expressed in closed form in terms of ${\cal M}$ and
${\cal Q}$.

What happens now if the system is at equilibrium? For this case there exists an energy functional $F$
(that can be either the entropy~\cite{deGroot}, a Lyapunov functional~\cite{Ojalvo}, the Hamiltonian,
 or a free energy~\cite{HH})
which is an integral over space of a local functional ${\cal F}$,
which depends on the field $\phi$ and its gradients.
The evolution equation for $\phi$ can be obtained by generalization to the continuum of the thermodynamics of
irreversible process (see, e.g., chapter~VII of~\cite{deGroot}). This theory relates the time evolution
of the fields with its conjugated variables $\Phi$, or the so-called thermodynamic forces, as
\begin{equation}\Label{Onsager}
\partial_t\phi = - {\cal L} \Phi+\xi({\bf r},t).
\end{equation}
Here ${\cal L}$ is the (symmetric) Onsager operator, which is a generalization to continuum of the
Onsager matrix of transport coefficients. It is also called the dissipation matrix.
The second law of thermodynamics requires that the real parts of the Onsager operator eigenvalues are positive in order to
guarantee local increase of entropy.
The fields $\Phi$ appearing in (\ref{Onsager}) are the conjugated variables
of $\phi$ and can be derived from the functional $F$ as
\begin{equation}\Label{OnsagerF}
\Phi=\frac{\delta F}{\delta \phi}.
\end{equation}

If the fluctuations are small and the system is far from a phase transition, we can assume that
$\delta F/\delta \phi$ is linear in the field $\phi$. This implies that $F[\phi]$ is bilinear in
the field $\phi$:
\begin{equation}
F[\phi] = \int d{\bf r}\, {\cal F}(\phi,\nabla \phi, \ldots) \equiv  \int d{\bf r} \,
\phi {\cal G} \phi,
\end{equation}
and the equation above defines the operator ${\cal G}$.
Although the definition of ${\cal F}$ may not be unique, ${\cal G}$ is unique~\cite{twoF},
and ${\cal G}$ must be positive definitive in order to guarantee the existence
of a minimum of the free energy. It can also be chosen to be Hermitian, because
the antisymmetric part does not contribute to the free energy.
In this way, the corresponding Langevin equation is linear in $\phi$. Therefore, we can write that
\begin{equation}\Label{defG}
\frac{\delta F}{\delta \phi} = {\cal G}\phi.
\end{equation}

Combination of Eqs.~(\ref{Onsager}) and (\ref{defG}) allows us to write the
evolution equation for $\phi$ as
\begin{equation}\Label{OnsagerFinal}
\partial_t\phi = -{\cal L}{\cal G}\phi+\xi({\bf r},t).
\end{equation}

If the system is at equilibrium, the well-known {\em fluctuation--dissipation theorem}~\cite{Kubo}
imposes that the intensity of the noise, given by ${\cal Q}$, must be related to
the Onsager operator ${\cal L}$ by
\begin{equation}
{\cal Q}=k_B T({\cal L}+{\cal L}^+). \label{FDtheorem.Q}
\end{equation}
Equation~(\ref{OnsagerFinal}) is formally equal to Eq.~(\ref{Langevin}), with the operator ${\cal M}$
 given by $\Operator{M}=\Operator{L}\Operator{G}$.
As both $\Operator{G}$ and $({\cal L}+{\cal L}^+)$ are Hermitian and definitive positive, it can be shown that the eigenvalues of
$\Operator{M}$ have positive real parts, even though $\Operator{M}$ can be
non-Hermitian or undefined (as in the case of the linear hydrodynamic equations)
(see~\cite{deGroot}, chapter~V)~\cite{HornJohnson}.
In both equilibrium and nonequilibrium dynamics we will assume that the real part of the spectrum of ${\cal M}$
is strictly positive, i.e., there are no neutral modes as happens when there is
continuous symmetry breaking~\cite{Forster} or critical phenomena~\cite{Critical1,Critical2}.

In equilibrium the fluctuation--dissipation relation has drastic consequences for the solution of the Langevin equation
associated to Eq.~(\ref{OnsagerFinal}). The equation (\ref{matrixg}) is now written
\begin{equation}\Label{matrixgEQ}
{\cal LGK}^{-1}+{\cal K}^{-1}{\cal GL^+} =k_BT  ({\cal L}+ {\cal L}^+),
\end{equation}
which admits the solution ${\cal K}= \beta {\cal G}$. Once substituted
into Eq.~(\ref{solutionFP}), the probability
distribution is given by the exponential of the functional $F$ multiplied by $\beta=(k_B T)^{-1}$.
More precisely,
\begin{equation}\Label{Peq}
P[\phi]=\frac{1}{Z} e^{ -\beta F[\phi]},
\end{equation}
where $Z$ is the partition function or the
normalization constant of $P$.
Given this probability distribution $P[\phi]$,
we can now calculate the average of any dynamical variable $A(\phi)$ as
\begin{equation}\Label{Faverage}
\langle A\rangle= \int d\phi A(\phi)  P[\phi].
\end{equation}
In particular, the average of the functional $F$
can be calculated as
\begin{equation}\Label{Faverage2}
\langle F\rangle= -\frac{\partial\ln Z}{\partial \beta}.
\end{equation}
Equations~(\ref{Peq}) and (\ref{Faverage2}) are only valid for equilibrium systems,
for which an energy functional exists and the fluctuation--dissipation theorem is valid.
However, if the system is out of equilibrium, the probability distribution is
not the exponential of $F$ and therefore its average is not given in terms of
the partition function $Z$.

\setcounter{equation}{0}
\section{Casimir forces from the average stress tensor}

How is this discussion related to the calculation of Casimir forces?
The Casimir force is normally calculated for equilibrium situations, that is,
when the noise is of thermal origin and the fluctuation--dissipation theorem is
satisfied.
One way to calculate the Casimir force is by evaluation of the stress
tensor ${\Tensor{T}}$.
From the functional $F$, the stress tensor is calculated as~\cite{LandauLifshitz}
\begin{align}\Label{stress}
\Tensor{T}_{ij}&= \Tensor{I}_{ij}{\cal F}-\nabla_i\phi
\frac{\partial {\cal F}}{\partial\nabla_j \phi}-
2\nabla_{ik}\phi \frac{\partial \cal F}{\partial\nabla_{kj} \phi}+\dots\nonumber \\
\Tensor{T}&\equiv  \TensorT[\phi,\phi,\Vec{r}],
\end{align}
which allows the definition of the symmetric bilinear stress tensor operator $\TensorT$.
For isotropic systems, the local stress is simply given by the diagonal components of the stress
tensor, or by one-third of its trace. Usual forms of $\TensorT$ are $\lambda\phi(\Vec{r})^2$ times the identity matrix or a tensorial product of gradient as in liquid crystals, but it can be also nonlocal, as in~\cite{Tarazona}.

Because of the intrinsically fluctuating nature of the fields, the stress tensor has
to be averaged over the random fields $\xi( {\bf r},t)$ or the probability distribution (\ref{solutionFP}).
Once we have the averaged stress tensor, the Casimir force over a body of surface $S$ is obtained as
\begin{equation}\Label{FCasimir}
\Vec{F}_C= -\oint_S \langle \Tensor{T}(\Vec{r})\rangle \cdot \, \hat{\bf n}\, dS,
\end{equation}
where the integral extends over the
surface of the embedded bodies and the vector $ \hat{\bf n}$ is a unit vector normal to the surface, pointing inward the body.

\begin{figure}[htb]
\includegraphics[width=\columnwidth]{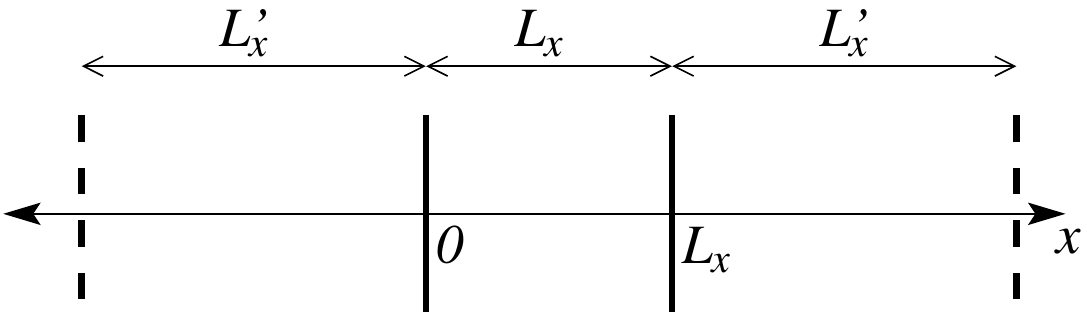}
\caption{Parallel-plate geometry used to compute the Casimir force. The system is confined between plates located at $x=0$ and $x=L_x$. Additional plates are located at distances $L_x'$ from these plates, and finally the limit $L_x'\to\infty$ is taken to mimic an infinite system.}
\label{fig.plates}
\end{figure}

As in the original Casimir calculation, the geometry that will be considered throughout this paper
consists of two parallel, infinite plates, perpendicular to the $x$-axis, separated by distance $L_x$ (Fig.~\ref{fig.plates}).
In this geometry, the Casimir force per unit area on the plates
is then the difference between the normal stress on the interior
and exterior side, where the latter is obtained by taking the limit $L'_x\to\infty$. 
The force per unit area on the left plate is
\beq\Label{FCA}
F_C/A = \left [\mean{\Tensor{T}_{xx}(x=0;L_x)} - \lim_{L'_x\to\infty} \mean{\Tensor{T}_{xx}(x=0;L'_x)} \right].
\eeq
The interpretation is that, if $F_C/A$ is negative, the plates repel each other, while if it is positive, an attraction between the plates appears.

Let us discuss Eq.~(\ref{FCasimir}) for equilibrium and nonequilibrium situations.
In the former case, i.e., a system in equilibrium, the average of the stress tensor
can be taken in two ways: as an average over the probability distribution
given by Eq.~(\ref{Peq}), or as an average over the fluctuating
term $\xi({\bf r},t)$. Equilibrium thermodynamics guarantees that both averages are the same.
In contrast, in a system out of equilibrium, we are left with one option,
the average over the noise $\xi({\bf r},t)$, because ${\cal K}$ cannot be obtained in general.
As mentioned, the system can be out of equilibrium if the fluctuation--dissipation relation is not satisfied.
In this case, there still exists a functional $F$ (from which the Langevin equation is constructed), and the
stress tensor can be defined via Eq.~(\ref{stress}). Then, the average in (\ref{FCasimir})
has to be taken over the noise.
Finally, a more complex situation is when the Langevin equation is in its most
general form, i.e., Eq.~(\ref{Langevin}), without ${\cal M}$ deriving from an Onsager matrix
and a functional $F$. In this case, the stress tensor cannot be constructed from Eq.~(\ref{stress}),
and one must appeal to other considerations in order to construct a stress tensor. One can use a microscopic
analysis of momentum transfer, kinetic theory, or invoke, for instance, the existence of a hydrostatic
pressure from which the Casimir force can be derived. We will assume, therefore, that it will always be possible
to build the stress tensor operator $\TensorT$.

\setcounter{equation}{0}
\section{Computation of Casimir forces}
In this section we will develop a formalism, valid for both equilibrium and nonequilibrium systems, that allows us to compute the average stress tensor and therefore the Casimir force. We will assume that the dynamics close to the stationary state is described by the dynamical equation (\ref{Langevin}), where the noise term is assumed to be Gaussian with vanishing mean. We assume that the noise has temporal and spatial correlations, but no cross-correlations,
\beq
\mean{\xi(\Vec{r}, t)\xi(\Vec{r}',t')} = h(\Vec{r}-\Vec{r}') c(t-t'). \Label{correlated.noise}
\eeq

Note that we have assumed a dynamical model whose deterministic part
is local in time (no memory) but that the noise can have some memory.
This possibility is not allowed by the fluctuation--dissipation theorem, and
therefore the system is automatically put out of equilibrium.
A necessary condition to recover equilibrium is a local correlation in time,
although this condition is not sufficient, as shown in Sect.~\ref{sec.2}.

To solve \reff{Langevin} we construct the left and right eigenvalue problems of ${\cal M}$
with the appropriate boundary conditions over the immersed bodies. Although we will consider the
case of two parallel plates, the formalism developed in this section is completely general.
The left and right eigenvalue problems read
\begin{eqnarray} \label{ev}
{\cal M}f_{n}\left(\textbf{r}\right) &=& \mu_{n}f_{n}\left(\textbf{r}\right),\\
{\cal M}^{+} g_{n}\left(\textbf{r}\right) &=& \mu_{n}^{*}g_{n}\left(\textbf{r}\right),
\end{eqnarray}
with the boundary conditions provided by ${\cal M}$
(which are the same as those of ${\cal L}$ if the dynamics derives from a free energy functional).
The left and right eigenfunctions are orthogonal under the scalar
product, i.e., $\mean{ g\vert f} = \int{d\textbf{r} g^{*}(\textbf{r}) f(\textbf{r}) }$;
that is, under appropriate normalization, $\mean{ g_n\vert f_m}=\delta_{nm}$.
We can project the field and the noise over the left eigenvalues
\begin{equation}\Label{functions-factoritation}
\phi(\textbf{r},t) = \sum_{n}\phi_{n}(t)f_{n}(\textbf{r}),
\quad
\xi(\textbf{r},t) = \sum_{n}\xi_{n}(t)f_{n}(\textbf{r}),
\end{equation}
where $\phi_{n}(t)=\mean{g_n\vert \phi(t)}$ and $\xi_{n}(t)=\mean{g_n\vert \xi(t)}$.
By inserting these expressions (\ref{functions-factoritation}) into the evolution equation (\ref{Langevin}) we get the evolution equation of each mode $\phi_{n}(t)$ as
\begin{equation}
\partial_{t}\phi_{n}(t) = -\mu_{n}\phi_{n}(t) + \xi_{n}(t).
\end{equation}
This equation can be solved analytically as
\begin{equation}\Label{temporal-mode-evolution}
\phi_{n}(t) = e^{-\mu_{n}t}\left[\phi_{n}(0) + \int_{0}^{t}e^{\mu_{n}\tau}\xi_{n}(\tau)d\tau \right].
\end{equation}
The first term $e^{-\mu_{n}t}\phi_{n}(0)$ is a transient term that vanishes for times
longer than $t\gg\frac{1}{{\rm Re}(\mu_{n})}$, so that
 the average of each mode over the noise $\xi$ is zero in this limit.

To compute the average stress tensor at each point, we need to compute $\mean{\TensorT[\phi,\phi,\Vec{r}]}$.
Expanding on the eigenvalue basis and using that $\phi=\phi^*$ we get that
\begin{eqnarray}
\langle\Tensor{T}(\Vec{r},t)\rangle &=& \sum_{m,n}\mean{\phi_n(t) \phi^*_m(t)} \TensorT_{nm}(\Vec{r}),
\end{eqnarray}
where $\TensorT_{nm}(\Vec{r})=\TensorT[f_n,f_m^*,\Vec{r}]$.

The cross-average of the mode amplitudes is obtained
from (\ref{temporal-mode-evolution}) and in the stationary regime
[$t\gg 1/{\rm Re}(\mu_n),1/{\rm Re}(\mu_m)$] can be written as
\begin{eqnarray}\Label{two-field-modes-average}
\mean{\phi_{n}(t)\phi_{m}^*(t)} &=& e^{ - (\mu_{n} + \mu_{m}^*)t}\int_{0}^{t} d\tau_{1}\int_{0}^{t}d\tau_{2}\nonumber\\
&& e^{\mu_{n}\tau_{1} + \mu_{m}^*\tau_{2}}
\mean{\xi_{n}(\tau_{1})\xi_{m}^*(\tau_{2})}.
\end{eqnarray}
Therefore, we need to calculate the correlation of the $n$ and $m$ components of the noise
\begin{equation}
\mean{\xi_{n}(\tau_{1})\xi_{m}^*(\tau_{2})} = \int\!\! d\textbf{r}_{1}\! \int\!\! d\textbf{r}_{2} g_{n}^*(\textbf{r}_{1})g_{m}(\textbf{r}_{2})\mean{\xi(\textbf{r}_{1},\tau_{1})\xi(\textbf{r}_{2},\tau_{2})}. \Label{correlation.noise.modes}
\end{equation}
Substituting Eq.~\reff{correlated.noise} into \reff{correlation.noise.modes}
and \reff{two-field-modes-average}, it is found that
\begin{equation} \Label{average-square-field}
 \lim_{t\to\infty}
\mean{ \phi_{n}(t)\phi_{m}^*(t)} = h_{nm}\frac{\widetilde{c}(\mu_{n})+\widetilde{c}(\mu^*_{m})}{\mu_{n} + \mu^*_{m}},
\end{equation}
where
\begin{equation}
h_{nm} = \int d\textbf{r}_1\int d\textbf{r}_{2}g^*_{n}(\textbf{r}_{1})h(\textbf{r}_{1} - \textbf{r}_{2})g_{m}(\textbf{r}_{2})=\mean{g_n\vert {\cal Q} g_m}
\end{equation}
and $\widetilde{c}$ is the Laplace transform of $c$.

Finally, the local average of the stress tensor in the stationary regime, where transients have been eliminated and the value of the stress tensor is independent of time, is given by
\begin{equation}
\langle\Tensor{T}(\Vec{r})\rangle = \sum_{nm} \frac{\widetilde{c}(\mu_{n})+\widetilde{c}(\mu^*_{m})}{\mu_{n} + \mu^*_{m}}
h_{nm} \TensorT_{nm}(\Vec{r}).
\end{equation}
This expression is generally divergent when summed over all eigenfunctions. This divergence
comes from the highest eigenvalues (corresponding to small wavelengths)
and is due to consider the mesoscopic dynamics given by Eq.~(\ref{Langevin}),
valid for all wavelengths.
However, it is only valid above a certain minimal distance (the atomic or molecular length, for example).
There are some techniques to avoid this divergence. For instance,
 a short-wavelength cutoff could be introduced as
in Ref.~\cite{BritoSotoPRE}, but here we will use regularization techniques similar to
the Riemann zeta function used in the electrodynamic case~\cite{Casimir Placas Paralelas}.

Using the previous expression, the conditions under which
Casimir forces exist in an equilibrium system can be deduced. As mentioned above, if the dynamics is local in time, the fluctuation--dissipation theorem implies that the noise terms must not have memory either, therefore $\widetilde{c}(\mu)=1/2$. Also, the equilibrium relation \reff{FDtheorem.Q} implies that $h_{nm} =k_B T (\mu_n+\mu_m^*)\mean{g_n\vert {\cal G}^{-1} g_m} $, and therefore the equilibrium average stress tensor simplifies to
\begin{equation}
\langle\Tensor{T}_{\rm eq}(\Vec{r})\rangle = k_B T\sum_{n,m} \mean{g_n\vert {\cal G}^{-1} g_m} \TensorT_{nm}(\Vec{r}).\label{Caso termico}
\end{equation}
If the free energy functional depends only on $\phi$ but not on its derivatives, the stress tensor operator turns out to be isotropic and is given by
$ \TensorT_{nm}(\Vec{r})=f_n\Operator{G}f_m^*\Tensor{I}_{3\times3}$ \reff{stress}. Then, thanks to the completeness of the basis, the stress tensor can be further simplified to $\langle\Tensor{T}_{\rm eq}(\Vec{r})\rangle =k_B T \delta(\Vec{r})$. This expression, once properly regularized, gives a stress that is independent of system size; that is, the stress is not renormalized by the fluctuations in a size-dependent way and therefore no Casimir force can be developed. On the contrary, if the stress tensor is not isotropic, as in the case of liquid crystals, the result is not trivial and Casimir forces can develop, as shown in~\cite{Casimir-liquidcrystals}.

All these equations provide expressions for the average fields and fluctuations, expressed in terms of the eigenvalues and eigenvectors of the problem, which encode the information of the evolution equation together with the boundary conditions.

To summarize, in this section we have proven that the Casimir force over a body is given by
\beq \Label{Force}
\Vec{F}_C = -\sum_{nm}  \frac{\widetilde{c}(\mu_{n})+\widetilde{c}(\mu^*_{m})}{\mu_{n} + \mu^*_{m}}h_{nm}
\oint_{S} \TensorT_{nm}(\Vec{r})\cdot \hat{{\bf n}}\, dS,
\eeq
which is the main result of this paper. It shows how to derive the Casimir force from the
dynamical equations for the field $\phi$ subjected to any kind of noise. It is obtained by diagonalizing the
evolution operator of the field, and projecting the noise correlation and the stress tensor over the set of eigenfunctions.
This approach provides the Casimir force for both equilibrium and nonequilibrium systems.

Equation~(\ref{Force}) shows the well-known nonadditive character of the Casimir force: neither the eigenvalues nor the eigenfunctions for different boundary conditions are easily related. They cannot be written as a sum of the eigenvalues and eigenfunctions of each different problem.

The rest of the paper deals with applications of Eq.~(\ref{Force}) to different physical systems, in both equilibrium and nonequilibrium situations.

\setcounter{equation}{0}
\section{Reaction--diffusion systems}
To show how this formalism works, we calculate the Casimir pressure between
two plane, infinite plates separated by distance $L_x$ immersed in a medium described by a quadratic free energy
\beq
F[\phi] = \int d{\bf r}\,  f_0 \phi^2(\Vec{r})/2.
\eeq
The multiplicative constant $f_0$ can be absorbed into $\phi$, and we will eliminate it in what follows.
The dynamics is described by two transport processes: relaxation and diffusion;
that is, the Onsager operator is ${\cal L} = \lambda - D\nabla^2$, where $\lambda$ and $D$ are the
transport coefficients (and consequently, positive) associated with the two irreversible processes of relaxation and diffusion, respectively.
The resulting equation is
\beq
\derpar{\phi}{t} = -\lambda \phi + D\nabla^2 \phi  +\xi(\Vec{r},t). \Label{eq.readdif}
\eeq

Fluctuation--dissipation is satisfied if the noise is delta correlated in time and the space correlation function is
\beq
h(\Vec{r}) = 2k_B T (\lambda -D \nabla^2) \delta(\Vec{r}).
\eeq

As the energy functional for this system is $\phi^2/2$,
without terms with spatial derivatives, the
stress tensor is identical to the local energy functional.
Also, the dynamic operator is Hermitian,
implying that eigenvalues are real and that there is no need to distinguish between left and right eigenfunctions.
In order to obtain Casimir forces the appropriate boundary conditions are of Neumann type,
as Dirichlet boundary condition would imply trivial vanishing forces.

We need to solve the eigenfunction problem for the spatial part of the
dynamics in order to calculate the average of the fields
that will lead to the Casimir pressure over the plates. So, we have to solve the
eigenfunction problem given by Eq.~(\ref{ev}) with ${\cal M}= \lambda - D\nabla^2$
obeying Neumann boundary conditions (no-flux boundary conditions), $\partial_{x}\phi(0,y,z) = \partial_{x}\phi(L_x,y,z) = 0$.

The normalized eigenfunctions are characterized by three indices $n_x$, $n_y$, and $n_z$, denoted as a whole by $n$, and their form is
\begin{eqnarray}\Label{autofunciones-Neumann}
& f_{n}(\textbf{r}) =  \sqrt{\frac{1}{V}}e^{i\textbf{k}_{\parallel}\cdot\textbf{r}_{\parallel}} &
\hspace{0.5cm}\text{if}\hspace{0.2cm}n_x=0\nonumber\\
& f_{n}(\textbf{r}) =  \sqrt{\frac{2}{V}}\cos\left(k_{x}x \right) e^{i\textbf{k}_{\parallel}\cdot\textbf{r}_{\parallel}} &
\hspace{0.5cm}\text{if}\hspace{0.2cm}n_x\geq 1.
\end{eqnarray}
Here, $\Vec{r}_{\parallel}=y\hat{\bf y}+z\hat{\bf z}$ and $\Vec{k}_{\parallel}=k_y\hat{{\bf y}}+k_z\hat{\bf z}$.
The eigenvalues are
\begin{equation}\Label{eigenvalues}
\mu_{n} = D\left(k_{x}^{2} + k_{y}^{2} + k_{z}^{2} + k_{0}^{2}\right),
\end{equation}
where $
k_{x}=\frac{\pi}{L_x}n_x$,
$k_{y}=\frac{2\pi}{L_{y}}n_{y}$, and
$k_{z}=\frac{2\pi}{L_{z}}n_{z}$,
with $n_{x}=0,1,2,\dots$ and $(n_{y},n_{z})\in\mathbb{Z}^{2}$. The quantity
$k_{0}^{-1}  =  \sqrt{D/\lambda}$ is the
characteristic correlation length of the system.

The average stress is then
\begin{equation}
\mean{\Tensor{T}_{xx}(\mathbf{r})} = \frac{1}{2D} \sum_{nm}
\frac{h_{nm}
\left[\widetilde{c}(\mu_n) + \widetilde{c} (\mu_m)\right]}{\Vec{k}_{n}^{2} + \Vec{k}_{m}^{2}  +2 k_{0}^{2}}f_{n}(\textbf{r})f^*_{m}(\textbf{r}). \Label{TensorTReacDif}
\end{equation}
This expression needs to be regularized, otherwise it is divergent.
The divergence, as explained in~\cite{BritoSotoPRE}, is due to the application of the mesoscopic model \reff{eq.readdif} up to very large wavevectors. Conceptually, the stress could be regularized by considering generalized hydrodynamic models valid for high wavevectors leading to finite stresses, but as the Casimir forces have their origin in the limitation of the fluctuation at small wavevectors, this is not necessary and other procedures are available. There are various regularization methods that allow the isolation of the divergent term that is independent of the plate separation and therefore cancels out in the computation of the Casimir force. The regularization method used in this manuscript is based on the Elizalde function detailed in the Appendix.

To obtain quantitative predictions, we consider specific cases for the noise correlations.

\subsection{Uncorrelated noise in time and space}
We first consider the case of a noise with vanishing correlation time and length, and intensity $\Gamma$, i.e.,
\beq
\mean{\xi({\bf r}, t) \xi({\bf r}',t')} = \Gamma \delta({\bf r}-{\bf r}')\delta(t-t'). \Label{doublewhitenoise}
\eeq
This noise correlation, without the $-\nabla^2\delta({\bf r})$ term,
automatically puts the system out of equilibrium, as ${\cal Q}\neq {\cal L}+{\cal L}^{+}$. The addition of such a term would have led to a stress that was independent of plate separation, not producing a Casimir force~\cite{CasimirGranular}. Therefore, we consider the effect of the nonequilibrium noise \reff{doublewhitenoise} on Casimir forces. In this case, $h_{nm}(\widetilde{c}(\mu_n)+\widetilde{c}(\mu_m))=\Gamma\delta_{nm}$ and the double sum in \reff{TensorTReacDif} is reduced.
On the surface of a plate and applying the limit $L_{y}, L_{z}\rightarrow\infty$, the stress is given by
\begin{align}
\mean{\Tensor{T}_{xx}(0)} &= \frac{\Gamma}{16\pi^{2}L_{x}D}\int_{-\infty}^{\infty}\!dk_{y}\int_{-\infty}^{\infty}\!dk_{z} \nonumber \\
& \sum_{n_x\in\mathbb{Z}}\frac{1}{\left(\frac{\pi n_x}{L_x}\right)^{2} + k_{y}^{2} + k_{z}^{2} + k_{0}^{2}} \nonumber\\
& = \frac{\Gamma}{8\pi L_{x}D} \int_{0}^{\infty}dkk\sum_{n_x\in\mathbb{Z}}\frac{1}{\left(\frac{\pi n_x}{L_x}\right)^{2} + k^{2} + k_{0}^{2}}, \Label{T.reacdif.whitenoise}
\end{align}
where polar coordinates in the $y$- and $z$-components have been used. Note that the original sum over $n_x$ in (\ref{TensorTReacDif}) runs only over $\mathbb{N}$, but the form of the normalizations of the eigenfunctions (\ref{autofunciones-Neumann}) allows extension of the sum over $\mathbb{Z}$ with a prefactor of $1/2$.

This expression is divergent, so it must be regularized. In
order to do so, we use the Chowla--Selberg expression shown in Eq.~(\ref{Chowla-Selberg 1d}).
The parameters are $s = 1$ , $\alpha=\pi/L_x$, and $\omega^2=k^2+k_0^2$.
The first term in the sum of (\ref{Chowla-Selberg 1d}) equals $L_x/\sqrt{k^2+k_0^2}$,
which combined with the prefactor in Eq.~(\ref{T.reacdif.whitenoise}) yields a term which is independent of $L_x$,
and therefore its contribution to the stress tensor cancels in virtue of Eq.~(\ref{FCA}). It must be remarked that the size-independent term is actually divergent if the continuous model is assumed to be valid for any wavevector.
Then, we are left with the infinite sum of modified Bessel functions $K_{1/2}$. This sum can be performed analytically,
with the result
\begin{align}
F_C/A & = \frac{\Gamma}{4\pi D}\int_{0}^{\infty}dk\frac{k}{\sqrt{k^{2} + k_{0}^{2}}} \frac{1}{e^{2\sqrt{k^{2} + k_{0}^{2}}L_x} - 1}\nonumber \\
    & = - \frac{\Gamma k_{0}}{8\pi D}\frac{\ln(1-e^{-2k_{0}L_x})}{ k_{0}L_x}. \label{reacc}
\end{align}
Let us note that, because the divergence was eliminated, we could have interchanged the integral with the summation of the modified Bessel functions
to obtain the same result.
Equation~(\ref{reacc}) shows that the Casimir force
diverges if the correlation length tends to infinity, i.e., if $k_0\to 0$.
This result was obtained in~\cite{BritoSotoPRE} using a regularizing kernel technique.

\subsection{Temporally correlated noise} \Label{temporalcorrelation.reacdif}

We next consider the case of a noise that is delta correlated in space but has exponential correlation in time
\begin{equation}
\mean{\xi\left(\textbf{r},t\right)\xi\left(\textbf{r}',t'\right)} = \Gamma \delta\left(\textbf{r} - \textbf{r}'\right)\left(1+\frac{a}{2}\right)e^{-a\abs{t - t'}}, \Label{noise.tempcorr}
\end{equation}
where the factor $(1+\frac{a}{2})$ allows both the white noise limit ($a\to\infty$) and the quenched noise limit ($a\to 0$) to be taken.
Again, the delta correlation in space leads to a term $\delta_{nm}$ that eliminates one summation in the stress at the plate, which is then given by
\begin{equation}
\mean{\Tensor{T}_{xx}(0)} = \frac{\left(1+\frac{a}{2}\right)\Gamma}{2V}\sum_{n}\frac{1}{\mu_{n} + a}\frac{1}{\mu_{n}}.
\end{equation}
If $a> 0$, we can factorize the quotient as
\begin{equation}
\mean{\Tensor{T}_{xx}(0)} = \frac{\left(1+\frac{a}{2}\right)\Gamma}{2aV}\sum_{n}\left[\frac{1}{\mu_{n}} - \frac{1}{\mu_{n} + a}\right],
\end{equation}
with $\mu_{n} = k_{x}^{2} + k_{y}^{2} + k_{z}^{2} + k_{0}^{2}$ as before. We note that this stress
is the difference between the Casimir stress of two systems with a white temporal noise \reff{T.reacdif.whitenoise} of
intensity $(1+a)\Gamma/a$, the first one with $k_{0}^{2} = \frac{\lambda}{D}$ and
the second one with $k_{1}^{2} = \frac{\lambda}{D} + \frac{a}{D}$. Then, the stress on the plate is
\begin{equation}\Label{aaa}
\mean{\Tensor{T}_{xx}(0)}  = -\frac{\Gamma\left(1+\frac{a}{2}\right)}{4a\pi DL_{x}}\ln\left(\frac{1 - e^{-2k_{0}L_x}}{1 - e^{-2k_{1}L_x}}\right).
\end{equation}
The Casimir force per unit surface on the plate is given just by this expression, because the stress
on the unbounded side vanishes [as shown by taking the limit $L_x\to\infty$ in Eq.~(\ref{aaa})].
Finally, we can reobtain the white noise limit if $a\rightarrow\infty$.

The case $a\to 0$ corresponds to the quenched limit, where static sources of noise are randomly distributed in space. The average normal stress at the wall is
\begin{equation}
\mean{\Tensor{T}_{xx}(0)} = \frac{\Gamma}{2V}\sum_{n}\frac{1}{\mu_{n}^{2}}.
\end{equation}
Taking the limit $L_y,L_z\to\infty$ and using polar coordinates,
\begin{align}
\mean{\Tensor{T}_{xx}(0)} = &\frac{\Gamma}{4\pi L_{x}D^{2}} \\
& \times \int_{0}^{\infty}dkk \sum_{n_{x}\in\mathbb{Z}}\left(\left(\frac{\pi n_{x}}{L_{x}}\right)^2 + k^{2} + k_{0}^{2} \right)^{-2} \nonumber.
\end{align}
Although this expression is finite and does not require a regularization procedure, the size-independent contribution can be eliminated using the same regularization procedure as before, using Eq.~(\ref{Chowla-Selberg 1d}) with $s = 2$. The result is
\begin{equation}
F_C/A = \frac{\Gamma}{4\pi D^{2}}\int_{0}^{\infty} dk\frac{k}{\omega^{3}}\frac{e^{2\omega L_x}\left(2\omega L_x  + 1\right) - 1}
{\left(e^{2\omega L_x} - 1\right)^{2}},
\end{equation}
where $\omega = \sqrt{k^{2} + k_{0}^{2}}$. After carrying out the integral, we obtain
\begin{equation}
F_C/A = \frac{\Gamma}{4\pi D^{2}k_{0}}\frac{1}{e^{2k_{0}L_{x}} - 1}.
\end{equation}
We remark that this system is not dynamically fluctuating, because the noise is quenched and the transients have been eliminated. Nevertheless, it creates a Casimir force whose origin is the same as previously considered in the sense that the presence of the second plate limits the spectrum of possible fluctuations, and therefore the renormalized stresses on the two sides of the plate are different.

\subsection{Maximally spatially correlated noise}
As a final case, we consider the situation in which the medium is perturbed externally by a spatially homogeneous noise, with vanishing correlation time. This could be the case when a rapidly fluctuating external field is applied to the medium.
\begin{equation}
\mean{\xi\left(\textbf{r},t\right)\xi\left(\textbf{r}',t'\right)} = \Gamma \delta\left(t - t'\right) \Label{maxspatialcorrnoise}.
\end{equation}
Applying the same computation procedure as in the other cases, the average local stress on each side of the plates is
\begin{equation}
\mean{\Tensor{T}_{xx}(0)}   = \frac{\Gamma}{2D}\frac{1}{2k_{0}^{2}} = \frac{\Gamma}{4\lambda},
\end{equation}
which is independent of the plate separation. Therefore, the Casimir force vanishes in this case.

\setcounter{equation}{0}
\section{Liquid crystals}
The existence of Casimir forces in liquid crystals has been known for some time now~\cite{Casimir-liquidcrystals}.
In this section we apply the formalism presented in Sect.~\ref{sec.2} to a nematic crystal, obtaining
the known Casimir force for an equilibrium situation, and expressions for the force for some
nonequilibrium conditions.
The free energy functional of a nematic liquid crystal~\cite{deGennes} can be written in terms of a planar field $\phi$ as
\beq
F=\int d{\bf r} \left[\frac{\kappa_1}{2} \phi^2 + \frac{\kappa_2}{2}(\nabla\phi)^2 \right], \Label{F.liqcry}
\eeq
where we have assumed that the director vector is written in terms
of the field $\phi$ as $\hat n=(\sin\phi,0,\cos\phi)$, together with
the one-constant approximation (proportional to $\kappa_2$).
The first term in Eq.~(\ref{F.liqcry}) comes from a magnetic field
directed along the $z$-axis, whose intensity is absorbed into $\kappa_1$.

The simplest dynamical model is obtained with a single relaxational transport coefficient, with the Onsager operator ${\cal L}=\lambda$, leading to
\beq
\derpar{\phi}{t} = -\lambda\kappa_1 \phi + \lambda\kappa_2 \nabla^2\phi + \xi, \Label{eq.liqcry}
\eeq
which is identical in form to \reff{eq.readdif}, but with three main differences: the form of the fluctuation--dissipation relation to be in equilibrium, the stress tensor, and the possible boundary conditions that produce Casimir forces.
Fluctuation--dissipation is realized, according to Sect.~\ref{sec.2}, if the noise satisfies
\beq
\mean{\xi({\bf r},t)\xi({\bf r}',t')} = 2k_B T \lambda \delta({\bf r}-{\bf r}')\delta(t-t'),
\eeq
i.e., is purely nonconservative.

Due to the presence of the gradient terms in the free energy
functional \reff{F.liqcry}, the stress tensor is not isotropic, and
therefore even in equilibrium Casimir forces can appear.
Using Eq.~(\ref{stress}) the $xx$ component of the stress tensor is
\beq
\Tensor{T}_{xx}= \frac{\kappa_1}{2}\phi^2 + \frac{\kappa_2}{2}\left(\derpar{\phi}{y}\right)^2 + \frac{\kappa_2}{2}\left(\derpar{\phi}{z}\right)^2 - \frac{\kappa_2}{2}\left(\derpar{\phi}{x}\right)^2.
\eeq
It is then possible to develop Casimir forces by imposing either Dirichlet or Neumann boundary conditions.
Dirichlet boundary conditions are equivalent to the strong anchoring conditions, i.e., $\phi=0$ over the surfaces, and will be the case studied here.
 Casimir forces with Neumann boundary conditions can be easily extracted from the Dirichlet ones.

In this case of Dirichlet boundary conditions, the eigenfunctions of the
operator ${\cal M}= \lambda\left[\kappa_1 -\kappa_2\nabla^2\right]$ are given by
$ f_{n}(\textbf{r}) = \sqrt{\frac{2}{V}}
\sin\left(k_{x}x \right) e^{i\textbf{k}_{\parallel}\cdot\textbf{r}_{\parallel}},$
with eigenvalues
\begin{equation}\Label{eigenvalues2}
\mu_{n} =\lambda\kappa_2( k_{x}^{2} + k_{y}^{2} + k_{z}^{2} + k_{0}^{2}),
\end{equation}
where $ k_{x}=\frac{\pi}{L_x}n_x$, $k_{y}=\frac{2\pi}{L_{y}}n_{y}$, $k_{z}=\frac{2\pi}{L_{z}}n_{z}$,
and $k_{0} = \sqrt{\frac{\kappa_1}{\kappa_2}}$,
 with indices $n_{x}\in\mathbb{N}$ and $(n_{y},n_{z})\in\mathbb{Z}^{2}$.

Because of the boundary conditions, the $xx$ component of the stress tensor at the plates is simply given by
\beq
\Tensor{T}_{xx}(0)= - \frac{\kappa_2}{2}\left(\derpar{\phi}{x}\right)^2.
\eeq

As in the case of the reaction--diffusion system, we will consider different types of noise correlations that, as will be shown below, produce Casimir forces of different character.

\subsection{Uncorrelated noise in time and space}
We consider an uncorrelated noise as described by \reff{doublewhitenoise}. This case can be considered as in equilibrium with a temperature given by $\Gamma = 2k_{B}T \lambda$. Again the double sum in Eq.~(\ref{Force}) can be reduced, and the stress tensor on the surface of a plate is given by
\begin{equation}
\mean{\Tensor{T}_{xx}(0)} = - \frac{\Gamma}{2V\lambda} \sum_{n}\frac{k_{x}^{2}}{k_{x}^{2} + k_{y}^{2} + k_{z}^{2} + k_{0}^{2}}.
\end{equation}
Applying the limit $L_{y}, L_{z}\rightarrow\infty$, we obtain
\begin{align}
\mean{\Tensor{T}_{xx}(0)} =&\frac{-\Gamma}{16 \pi^{2}L_{x}\lambda} \int_{-\infty}^{\infty}\!dk_{y}\int_{-\infty}^{\infty}\!dk_{z}\nonumber\\
& \sum_{n_{x}\in\mathbb{Z}}\frac{\left(\frac{\pi n_{x}}{L_{x}}\right)^{2}}{\left(\frac{\pi n_{x}}{L_{x}}\right)^{2} + k_{y}^{2} + k_{z}^{2} + k_{0}^{2}}.
\end{align}

Using polar coordinates and regularizing the resulting expression using Eqs.~(\ref{Relacion de Y con Z 2}) and (\ref{Chowla-Selberg 1d}) with $s = 1$,
\begin{align}
F_C/A =& \frac{\Gamma}{4\pi\lambda}\int_{0}^{\infty}dkk\frac{\sqrt{k^{2} + k_{0}^{2}}}{e^{2\sqrt{k^{2} + k_{0}^{2}}L_x} - 1}\\
=& \frac{\Gamma}{16\pi\lambda L_x^{3}}\left[\Li_{3}(e^{-2k_{0}L_x}) + 2k_{0}L_x\Li_{2}(e^{-2k_{0}L_x}) \right. \nonumber \\
& \left. + 2k_{0}^{2}L_x^{2}\Li_{1}(e^{-2k_{0}L_x})\right],
\end{align}
where $\Li_{s}(z) = \sum_{n=1}^{\infty}\frac{z^{n}}{n^{s}}$ is the polylogarithm function. At distances long compared with the correlation lenght, that is $L_{x}\gg k_{0}^{-1}$, the force decays as
\begin{equation}
F_C/A = \frac{\Gamma k_{0}^{2}}{8\pi\lambda L_{x}} e^{-2 k_{0}L_{x}}.
\end{equation}
In the opposite limit, when the plates are at a distance much smaller than the correlation lenght, or $L_{x}\ll k_{0}^{-1}$, the force is
\begin{equation}
F_C/A = \frac{\Gamma}{16\pi\lambda}\frac{\zeta(3)}{ L_{x}^{3}}. \label{fza.cliq.white}
\end{equation}
This result has already been obtained in the context of liquid crystals in~\cite{Ajdari} if we use the
fluctuation--dissipation theorem. It is also the high-temperature limit of the electromagnetic Casimir force between two perfect metal plates~\cite{Review Casimir}
(described by Dirichlet boundary conditions),
and by using the classical limit of the fluctuation--dissipation theorem.

\subsection{Temporally correlated noise}
We consider the temporally correlated noise described in Eq.~(\ref{noise.tempcorr}). By
using the eigenfunctions of the Dirichlet problem, the stress tensor over a plate takes the value
\begin{equation}
\mean{\Tensor{T}_{xx}(0)} = - \frac{\kappa_{2}\Gamma\left(1 + \frac{a}{2}\right)}{V}\sum_{n}
\frac{1}{\mu_{n} + a}\frac{k_{x}^{2}}{\mu_{n}}.
\end{equation}
For any $a\neq 0$, the same factorization method as used in Sect.~\ref{temporalcorrelation.reacdif} can be used, leading to a Casimir force per unit surface equal to
\begin{align}
F_C/A = & 
\frac{\left(1 + \frac{a}{2}\right)\Gamma}{8\pi a\lambda L_{x}^{3}}\left[\Li_{3}(e^{-2k_{0}L_x}) + 2k_{0}L_x\Li_{2}(e^{-2k_{0}L_x})\right. \nonumber\\
 & + 2k_{0}^{2}L_x^2\Li_{1}(e^{-2k_{0}L_x}) - \Li_{3}(e^{-2k_{1}L_x}) \nonumber\\
 &\left.- 2k_{1}L_x\Li_{2}(e^{-2k_{1}L_x})  - 2k_{1}^{2}L_x^2\Li_{1}(e^{-2k_{1}L_x}) \right],
\end{align}
where $k_{0}^{2} = \frac{\kappa_{1}}{\kappa_{2}}$ and $k_{1}^{2} = \frac{\kappa_{1}}{\kappa_{2}} + \frac{a}{\lambda\kappa_{2}}$.
In the limit of infinite correlation length we have $k_{0}\rightarrow 0$ and $k_{1}\rightarrow\sqrt{\frac{a}{\lambda\kappa_{2}}}$, from which we obtain
\begin{eqnarray}
F_C/A &=& 
\frac{\left(1 + \frac{a}{2}\right)\Gamma}{8\pi a\lambda L_{x}^{3}}
\left[\zeta(3) - \Li_{3}(e^{-2 k_{1}L_{x}}) \right. \\
&& \left. - 2k_{1}L_{x}\Li_{2}(e^{-2k_{1}L_{x}}) - 2k_{1}^{2}L_{x}^{2}\Li_{1}(e^{-2k_{1}L_{x}})\right].\nonumber
\end{eqnarray}
The presented result should be compared with~\cite{Bartolo}, where the same system was studied, but a different answer was given. At long distances it decays as in the case of white noise (\ref{fza.cliq.white}) with a prefactor $(1 + \frac{a}{2})/a$.

For $a\to 0$, we obtain the quenched limit of the stress tensor at the plates
\begin{equation}
\mean{\Tensor{T}_{xx}(0)}  = \frac{-\Gamma}{4\pi\lambda^{2}\kappa_{2}L_x}
\int_{0}^{\infty}dkk \sum_{n_{x}\in\mathbb{Z}}\frac{\left(\frac{n_{x}\pi}{L_{x}}\right)^2}{\left(\left(\frac{n_{x}\pi}{L_{x}}\right)^2 + k^{2} + k_{0}^{2} \right)^{2}}.
\end{equation}
This expression is regularized using Eqs.~(\ref{Relacion de Y con Z 2}) and (\ref{Chowla-Selberg 1d}) with $s = 2$, resulting in
\begin{equation}
F_C/A = \frac{\Gamma}{4\pi\lambda^{2}\kappa_{2}}\int_{0}^{\infty} dk\frac{k}{\omega}\frac{\left(1 - e^{2\omega L_x} + 2\omega L_x
e^{2\omega L_x}\right)}{\left(e^{2\omega L_x} - 1\right)^{2}},
\end{equation}
where $\omega = \sqrt{k^{2} + k_{0}^{2}}$. This integral can be carried out to obtain the Casimir force as
\begin{equation}
F_C/A = \frac{\Gamma}{4\pi\lambda^{2}\kappa_{2}L_{x}}\frac{k_{0}L_{x}}{e^{2k_{0}L_{x}} - 1}.
\end{equation}
In the limit of infinite correlation length we have
\begin{equation}
F_C/A = \frac{\Gamma}{8\pi\lambda^{2}\kappa_{2}L_{x}},
\end{equation}
and in the limit of small correlation length ($k_{0}L_{x}\gg 1$) we obtain
\begin{equation}
F_C/A = \frac{\Gamma k_{0}}{4\pi\lambda^{2}\kappa_{2}}e^{-2k_{0}L_{x}}.
\end{equation}

\subsection{Maximally spatially correlated noise}
As a final case we consider a noise that is rapidly fluctuating in time but that is homogeneous in space, described by the correlation \reff{maxspatialcorrnoise}. In this case, $h_{nm}$ is not diagonal but is given by
\begin{equation}
h_{nm} = 2V\Gamma \frac{[1 - (-1)^{n_{x}}][1 - (-1)^{m_{x}}]}{\pi^2 n_x m_x}\delta_{n_{y}0}\delta_{n_{z}0}\delta_{m_{y}0}\delta_{m_{z}0}.
\end{equation}

The stress on the plates is then given by
\begin{align}
\mean{ \Tensor{T}_{xx}(0)}
&  = - \frac{2\Gamma}{\lambda\pi^{2}}\sum_{n_{x},m_{x}=1}^{\infty}\frac{[1 - (-1)^{n_{x}}][1 - (-1)^{m_{x}}]}{n_{x}^{2} + m_{x}^{2} + 2\left(\frac{k_{0} L_x}{\pi}\right)^{2}}.
\end{align}

As in the double summation above only odd values of $n$ and $m$ are summed, it can be
expressed in terms of the
Elizalde zeta function over odd numbers, defined as
\begin{equation}
Z_{I}(\alpha ,\beta ,\omega ,s) = \sum_{n,m\in\mathbb{Z}}\frac{1}{\left(\alpha^{2}(2n + 1)^{2} + \beta^{2}(2m + 1)^{2} + \omega^{2}\right)^{s}},
\end{equation}
which can be written in terms  of four Elizalde zeta functions.
Using the asymptotic expansion of the Elizalde zeta functions given in Eq.~(\ref{Eliz}) with $p=2$ and $s=1$, the Casimir force can be expressed as an infinite sum of Bessel functions $K_0(x) $ with different values of $x$. The divergent terms, given by the first term in Eq.~(\ref{Eliz}), are independent of $L_x$, so the final expression is finite and given by
\begin{equation}
F_C/A = - \frac{\Gamma}{\lambda}\sum_{n\in\mathbb{Z}}\sum_{m\in\mathbb{Z}}^{\hspace{-0.6cm}(n,m)\neq(0,0)}\left[ \begin{array}{l}
 \phantom{+}K_{0}\left(2\sqrt{2}k_{0}L_x\sqrt{n^{2} + m^{2}}\right)\\
 - \frac{1}{2}K_{0}\left(2\sqrt{2}k_{0}L_x\sqrt{\frac{n^{2}}{4} + m^{2}}\right)\\
 - \frac{1}{2}K_{0}\left(2\sqrt{2}k_{0}L_x\sqrt{n^{2} + \frac{m^{2}}{4}}\right)\\
 + \frac{1}{4} K_{0}\left(2\sqrt{2}k_{0}L_x\sqrt{\frac{n^{2}}{4} + \frac{m^{2}}{4}}\right)
\end{array} \right].
\end{equation}

Two limiting cases can be considered to clarify this result. First, in the limit of long distances $k_{0}L_x\gg 1$, the Casimir force is given by
\begin{equation}
F_C/A = 2\frac{\Gamma}{\lambda}\sqrt{\frac{\sqrt{2}\pi}{k_{0}L}}e^{-\sqrt{2}k_{0}L},
\end{equation}
whereas in the opposite limit of long correlation length $k_0L\ll 1$, the result is
\begin{equation}
F_C/A = - \frac{\Gamma}{\lambda}\log\left(k_{0}L\right).
\end{equation}
In the limit of infinite correlation length this result diverges.

\setcounter{equation}{0}
\section{Two-field system}
In the two systems we have considered so far (reaction--diffusion and liquid crystals), the dynamics is described by a Hermitian operator and therefore the potential of the method described herein is not fully evident. In this section, we build a more complex system, described by a model with two fields (which could be temperature and concentration, for example) coupled in a nonsymmetric way. For simplicity and to be concrete we will consider that the fields $\psi_1$ and $\psi_2$ are scalar,  subject to Neumann boundary conditions, and with eigenfunctions described by the Fourier modes \reff{autofunciones-Neumann}. In Fourier space, the dynamic equation is
\beq
\derpar{}{t}  \begin{pmatrix} \psi_1\\ \psi_2 \end{pmatrix} =
- \begin{pmatrix} \alphak & 0 \\ \betak & \gammak \end{pmatrix}
\begin{pmatrix} \psi_1\\ \psi_2 \end{pmatrix}
+ \begin{pmatrix} \xi_1\\ \xi_2 \end{pmatrix}.
\eeq
The noises are assumed to be white with different correlation intensities (allowing one of them to be set equal to zero later) and no cross-correlation
\begin{align}
\mean{\xi_1({\bf r}, t) \xi_1({\bf r}',t')} &= \Gamma_1 \delta({\bf r}-{\bf r}')\delta(t-t'),\nonumber \\
\mean{\xi_2({\bf r}, t) \xi_2({\bf r}',t')} &= \Gamma_2 \delta({\bf r}-{\bf r}')\delta(t-t'),\nonumber \\
\mean{\xi_1({\bf r}, t) \xi_2({\bf r}',t')} &= 0.
\end{align}

Finally, the stress tensor is assumed to be isotropic, depending only on the fields as
\beq
\Tensor{T}_{xx} = \kappa_1 \psi_1^2 + \kappa_2 \psi_2^2 .
\eeq

As the dynamic matrix is non-Hermitian, the left and right eigenmodes are different, being given by
\begin{align}
f_{1,\Vec{k}}(\textbf{r}) &= \sqrt{\frac{2}{V}}\cos\left(k_{x}x \right) e^{i\textbf{k}_{\parallel}\cdot\textbf{r}_{\parallel}} \begin{pmatrix} 1 \\ \frac{\betak}{\alphak-\gammak} \end{pmatrix},\nonumber \\
f_{2,\Vec{k}}(\textbf{r}) &= \sqrt{\frac{2}{V}}\cos\left(k_{x}x \right) e^{i\textbf{k}_{\parallel}\cdot\textbf{r}_{\parallel}} \begin{pmatrix} 0 \\ 1 \end{pmatrix},\nonumber \\
g_{1,\Vec{k}}(\textbf{r}) &= \sqrt{\frac{2}{V}}\cos\left(k_{x}x \right) e^{i\textbf{k}_{\parallel}\cdot\textbf{r}_{\parallel}} \begin{pmatrix} 1 \\ 0 \end{pmatrix},\nonumber \\
g_{2,\Vec{k}}(\textbf{r}) &= \sqrt{\frac{2}{V}}\cos\left(k_{x}x \right) e^{i\textbf{k}_{\parallel}\cdot\textbf{r}_{\parallel}} \begin{pmatrix} -\frac{\betak}{\alphak-\gammak} \\ 1, \end{pmatrix}
\end{align}
with eigenvalues
\begin{align}
\mu_{1,\Vec{k}} &= \alphak,\nonumber \\
\mu_{2,\Vec{k}} &= \gammak.
\end{align}

Using these eigenmodes, the different elements needed to compute the Casimir pressure are
\begin{align}
h_{i\Vec{k},j\Vec{q}} &= \delta_{\Vec{k}\Vec{q}}
\begin{pmatrix}
\Gamma_1 & - \Gamma_1 \frac{\betak}{\alphak-\gammak} \\
- \Gamma_1 \frac{\betak}{\alphak-\gammak} & \Gamma_2- \Gamma_1^2  \frac{\betak^2}{(\alphak-\gammak)^2}
\end{pmatrix}, \nonumber \\
T_{i\Vec{k},j\Vec{q}} &= \delta_{\Vec{k}\Vec{q}}
\begin{pmatrix}
\kappa_1+\kappa_2  \frac{\betak^2}{(\alphak-\gammak)^2} & \kappa_2  \frac{\betak}{\alphak-\gammak}\\
 \kappa_2  \frac{\betak}{\alphak-\gammak} & \kappa_2
\end{pmatrix} .
\end{align}

After simple algebra, the stress tensor on the plates is obtained as
\begin{align}
\mean{\Tensor{T}_{xx}(0)} = \frac{2}{V} \sum_{\Vec{k}} &\left[
\frac{\Gamma_1(\kappa_1+\kappa_2\betak^2/(\alphak-\gammak)^2}{2\alphak} \right. \nonumber \\
& \frac{2\Gamma_1 \kappa_2 \betak^2/(\alphak-\gammak)^2}{\alphak+\gammak}  \nonumber \\
&  \left.- \frac{(\Gamma_1 \betak^2/(\alphak-\gammak)^2 +\Gamma_2) \kappa_2}{2\gammak}
\right],
\end{align}
which for specific models (that is, specific values of $\alphak$, $\betak$, and $\gammak$) could be computed and regularized to obtain the Casimir force on the plates.

To show the kind of results that can be obtained we consider the simple reaction--diffusion two-field model $\alphak=\lambda_1+Dk^2$, $\betak=\lambda_{12}$, and $\gammak=\lambda_2+Dk^2$, with noise intensities $\Gamma_1=\Gamma$ and $\Gamma_2=0$~\cite{warningjordan}, representing the system
\begin{align}
\derpar{\psi_1}{t} &= -\lambda_1\psi_1 + D\nabla^2 \psi_1 +\xi, \nonumber \\
\derpar{\psi_2}{t} &=  -\lambda_2\psi_2 + D\nabla^2 \psi_2 -\lambda_{12}\psi_1 .
\end{align}

Furthermore, we assume that the stress only depends on $\psi_2$, i.e., $\Tensor{T}_{xx}=\kappa\psi_2^2$. Therefore, any eventual Casimir force is produced by the fluctuations of the second field which are produced by the coupling with the first field. The stress on the plates is finally
\beq
\mean{\Tensor{T}_{xx}(0)} = \frac{\Gamma\kappa\lambda_{12}^2}{2V} \sum_{\Vec{k}} \left[(\lambda_1+Dk^2)(\lambda_2+Dk^2)(\lambda_1+\lambda_2+2Dk^2)\right]^{-1}. \nonumber
\eeq
Assuming that $\lambda_{1}\neq\lambda_{2}$, we can apply partial fraction decomposition to obtain
\begin{equation}
\mean{\Tensor{T}_{xx}(0)} = \frac{\Gamma\kappa\lambda_{12}^2}{2D V(\lambda_{1} - \lambda_{2})^{2}} \sum_{\Vec{k}}\left[\frac{1}{k^{2} + k_{1}^{2}} + \frac{1}{k^{2} + k_{2}^{2}} - \frac{2}{k^{2} + k_{3}^{2}}\right], \nonumber
\end{equation}
where $k_{1}^{2} = \lambda_{1}/D$, $k_{2}^{2} = \lambda_{2}/D$, and $k_{3}^{2} = (\lambda_{1} + \lambda_{2})/2D$. Then, we can perform each infinite sum as in the case of scalar white noise to obtain the Casimir force as
\begin{equation}
\mean{\Tensor{T}_{xx}(0)} = \frac{- \Gamma\kappa\lambda_{12}^{2}}{4\pi DL_{x}(\lambda_{1} - \lambda_{2})^{2}}\ln\left(\frac{\left(1 - e^{-2k_{1}L_x}\right)\left(1 - e^{-2k_{2}L_x}\right)}{\left(1 - e^{-2k_{3}L_x}\right)^{2}}\right). \nonumber
\end{equation}
It is interesting to note that, if $\lambda_{1} = 0$ and/or $\lambda_{2} = 0$, this Casimir force diverges.

\section*{Conclusions}
In this article, we have developed a formalism to study Casimir forces in classical systems out of equilibrium based on the stochastic dynamical equations of the system under study. The equilibrium case is recovered as a particular limit where the fluctuation--dissipation theorem is valid.

In particular, we study the interaction which appears between intrusions in a medium subject to any kind of noise. The study is restricted to additive noise and nonquantum systems; quantum cases and multiplicative noises are left for future work. The method only relies on the stochastic evolution equation of the field in the medium, and information about the interaction between the medium and the intrusions, as given by the boundary conditions of the fields at the surface of the bodies. No assumptions are made regarding any characteristic of the noise, which could be internal of external, thermal or induced, white or colored, and even non-Gaussian.

This formalism reduces to the classical thermal Casimir effect when the medium is subjected to an additive Gaussian white noise with autocorrelation amplitude $\mathcal{Q} = k_{B}T(\mathcal{L}+\mathcal{L}^{+})$ and its dynamics is described by a Hermitian operator, as shown in \eqref{Caso termico}.

We have obtained an exact formula for the Casimir force felt by a body \eqref{Force}, which shows how to derive the Casimir force for any geometrical configuration and noise. Equation \eqref{Force} can be used to obtain non-equilibirum induced self forces over asymmetric bodies, as shown in \cite{Buenzli-Soto}. It can also provide a numerical tool useful to evaluate
Casimir forces for complicated geometries. 
 Along this paper, we have used the  formula to obtain the force between parallel plates in different media (in a reaction--diffusion model and in liquid crystals) under the influence of different kinds of Gaussian noises, i.e., white noise to recover the thermal case already studied in the literature, and noises with nonzero spatial or temporal correlation lengths, where different forces appears.

Finally we have shown an example of the evaluation of Casimir forces in a system with non-Hermitian evolution dynamics, which was an intractable problem until the development of the formalism presented herein.

\section*{Appendix: Elizalde zeta function}
\setcounter{equation}{0}
\renewcommand{\theequation}{A.\arabic{equation}}

The computation of the Casimir forces makes use of the asymptotic expansion of the Elizalde zeta function, which is defined as~\cite{Elizalde}
\begin{equation}
Z_{p}(s,a_{i},\omega) = \sum_{\Vec{n}\in\mathbb{Z}^{p}}\frac{1}{\left(\sum_{i=1}^{p}a_{i}^{2}n_{i}^{2} + \omega^{2}\right)^{s}}
\end{equation}
and admits the asymptotic expansion, valid for all complex $s$,
\begin{eqnarray}\label{Eliz}
Z_{p}(s,a_{i},\omega) & = & \frac{\Gamma(\frac{p}{2})\Gamma(s - \frac{p}{2})}{\Gamma(s)\prod_{i=1}^{p}a_{i}}\omega^{p - 2s}\nonumber\\
& + & \frac{2\pi^{s}\omega^{\frac{p}{2} - s}}{\Gamma(s)\prod_{i=1}^{p}a_{i}} \sum_{\Vec{n}\in\mathbb{Z}^{p} - \{\vec{0}\}}\left(\sum_{i=1}^{p}\left(\frac{n_{i}}{a_{i}}\right)^{2}\right)^{\frac{2s - p}{4}} \nonumber\\
&& K_{s - \frac{p}{2}}\left(2\pi\omega\sqrt{\sum_{i=1}^{p}\left(\frac{n_{i}}{a_{i}}\right)^{2}}\right),
\end{eqnarray}
where $K_\nu (z)$ is the inhomogeneous Bessel
function of the second kind or Macdonald function. When $p = 2$, this is called
the Chowla--Selberg formula. We are mainly interested in
the case $p = 1$ with $\alpha > 0$, for which
\begin{align}\Label{Chowla-Selberg 1d}
Z_{1}(s,\alpha,\omega) = & \sum_{n\in\mathbb{Z}}\frac{1}{\left(\alpha^{2}n^{2} + \omega^{2}\right)^{s}}\nonumber\\
                        = & \frac{\Gamma(\frac{1}{2})\Gamma(s - \frac{1}{2})}{\Gamma(s)\alpha}\omega^{1 - 2s}  \nonumber \\
                       & + \frac{4\pi^{s}}{\Gamma(s)\alpha} \sum_{n = 1}^{\infty}\left(\frac{n}{\alpha\omega}\right)^{s - \frac{1}{2}}K_{s - \frac{1}{2}}\left(2\pi\omega\frac{n}{\alpha}\right).
\end{align}

When studying Casimir forces between plates with Dirichlet boundary conditions, the following series needs to be computed
\begin{equation}
Y_{1}(s,\alpha,\omega) = \sum_{n\in\mathbb{Z}}\frac{\alpha^{2}n^{2}}{\left(\alpha^{2}n^{2} + \omega^{2}\right)^{s}}.
\end{equation}
It is straightforward  to obtain
\begin{align}\Label{Relacion de Y con Z}
\sum_{n\in\mathbb{Z}}\frac{\alpha^{2}n^{2}}{\left(\alpha^{2}n^{2} + \omega^{2}\right)^{s}} = &
\sum_{n\in\mathbb{Z}}\frac{1}{\left(\alpha^{2}n^{2} + \omega^{2}\right)^{s-1}} \nonumber \\
&- \omega^{2}\sum_{n\in\mathbb{Z}}\frac{1}{\left(\alpha^{2}n^{2} + \omega^{2}\right)^{s}},
\end{align}
and therefore
\begin{equation}\Label{Relacion de Y con Z 2}
Y_{1}(s,\alpha,\omega) = Z_{1}(s - 1,\alpha,\omega) - \omega^{2}Z_{1}(s,\alpha,\omega).
\end{equation}

\acknowledgments{
This article has benefitted from discussions with many colleagues:
J.\,M.\,R.~Parrondo, M.~Clerc, N.~van Kampen, O.~Descalzi, F.~Barra,
A.~Galindo, G.\,G.~Alcaine, J.~San Mart\'{\i }n and U.M.B. Marconi.
P.R.-L. and R.B. are supported by the Spanish projects
MOSAICO, UCM/PR34/07-15859, and MODELICO (Comunidad de Madrid).
 P.R.-L.'s research is also supported by a FPU MEC grant.
The research is supported by Fondecyt grants 1100100, 1070958, and 7070301, 
and {\em Proyecto Anillo  ACT 127}. }


\section*{References}
\begin{thebibliography}{10}
\bibitem{KardarGolestanian} M. Kardar and R. Golestanian, Rev. Mod. Phys. \textbf{71}, 1233 (1999).
\bibitem{Casimir Placas Paralelas} H. B. G. Casimir, Proc. K. Ned. Akad. Wet. \textbf{51}, 793 (1948).
\bibitem{Kardar-Geometrias-Arbitrarias} T. Emig, N. Graham, R.L. Jaffe, and M. Kardar, Phys. Rev. Lett. \textbf{99}, 170403 (2007). 
\bibitem{Cardy} J. L. Cardy in \textit{Introduction to Phase Transitions and Critical Phenomena}, edited by C. Domb and J. L. Lebowitz (Academic, New York, 1987) [Vol. 11; ISBN-13: 978-0195053166].
\bibitem{Bartolo} D. Bartolo, A. Ajdari, and J. B. Fournier, Phys. Rev. E. \textbf{67}, 061112 (2003).
\bibitem{Krech} M. Krech. Phys, Rev. E. \textbf{56}, 1642 (1997).
\bibitem{Dean1} D. S. Dean and A. Gopinathan, J. Stat. Mech. L08001 (2009).
\bibitem{Ajdari} A. Ajdari, B. Duplantier, D. Hone, L. Pelity, and J. Prost, J. Phys. II France \textbf{2}, 487 (1992).
\bibitem{CasimirGranular} R. Brito, R. Soto, and U. Marini Bettolo Marconi, Gran. Matt. \textbf{10}, 29 (2007).
\bibitem{Dean2} D. S. Dean and A. Gopinathan, Phys. Rev. E \textbf{81}, 041126 (2010).
\bibitem{Sagues} F. Sagu\'es, J. M. Sancho, and J. Garc\'{\i}a-Ojalvo,  Rev. Mod. Phys. \textbf{79}, 829 (2007).
\bibitem{Gollub} J. P. Gollub and J. F. Steinman, Phys. Rev. Lett. \textbf{45}, 551 (1980).
\bibitem{Cattuto} C. Cattuto, R. Brito, U. Marini Bettolo Marconi, F. Nori, and R. Soto, Phys. Rev. Lett. \textbf{96}, 178001 (2006).
\bibitem{Brandt} H. R. Brand, S. Kai, and S. Wakabayashi, Phys. Rev. Lett. \textbf{54}, 555 (1985).
\bibitem{photosensitive} L. Kuhnert, K. I. Agladze, and V. I. Krinsky, Nature \textbf{337}, 244 (1989).
\bibitem{NajafiGolestanian} A. Najafi and R. Golestanian, EPL \textbf{68}, 776 (2004).
\bibitem{Gardiner} C. Gardiner, \textit{Handbook of Stochastic Methods} (Springer Series in Synergetics, 2004) [ISBN-13: 978-3540208822].
\bibitem{Zwanzig} R. Zwanzig, \textit{Nonequilibrium Statistical Mechanics} (Oxford University Press, 2001) [ISBN-13: 978-0195140187].
\bibitem{deGroot} S. R. de Groot and P. Mazur, \textit{Non-Equilibrium Thermodynamics} (Dover, 1984) [ISBN-13: 978-0486647418].
\bibitem{vanKampen} N. G. van Kampen, \textit{Stochastic Processes in Physics and Chemistry} (North Holland, 2007) [ISBN-13: 978-0444529657].
\bibitem{Risken} H. Risken, \textit{The Fokker-Planck Equation: Methods of Solutions and Applications} (Springer Series in Synergetics, 1996) [ISBN-13: 978-3540615309].
\bibitem{Ojalvo} J. Garc\'{\i}a-Ojalvo and J. M. Sancho, \textit{Noise in Spatially extended systems} (Springer, 1999) [ISBN-13: 978-0387988559].
\bibitem{HH} P. C. Hohenberg and B. I. Halperin, Rev. Mod. Phys. \textbf{49}, 435 (1977).
\bibitem{twoF}
For instance the two local energy functions ${\cal F}_1 = |\nabla \phi|^2/2$ and
${\cal F}_2 = -\phi \Delta  \phi/2$ yield the same energy functional $F$
and the same ${\cal G}$ operator: ${\cal G}=\Delta$.
\bibitem{Kubo} R. Kubo, M. Toda, and N. Hashitsume, \textit{Statistical Physics II} (Springer Verlag) [ISBN: 3-540-11461-0].
\bibitem{HornJohnson} R. A. Horn and C. R. Johnson, \textit{Topics in Matrix Analysis}, theorem 2.4.15 (Cambridge University Press, 1994).
\bibitem{Forster} D. Forster, \textit{Hydrodynamic Fluctuations, Broken Symmetry, and Correlation Functions} (HarperCollins Canada, 1994).
\bibitem{Critical1} O. Descalzi and R. Graham, Phys. Lett. A \textbf{170}, 84 (1992).
\bibitem{Critical2} A. Macio\l{}ek, A. Gambassi, and S. Dietrich, Phys. Rev. E \textbf{76}, 031124 (2007).
\bibitem{Tarazona} R. Evans and P. Tarazona, Phys. Rev. Lett. \textbf{52}, 557 (1984).
P. C. Hemmer and J. L. Lebowitz, in \textit{Phase Transitions and Critical Phenomena}, edited by C. Domb and M. S. Green (Academic, New York, 1976) [Vol. VB].
\bibitem{Casimir-liquidcrystals} A. Ajdari, L. Peliti, and J. Prost, Phys. Rev. Lett. \textbf{66}, 1481 (1991).
\bibitem{BritoSotoPRE} R. Brito, U. M. B. Marconi, and R. Soto, Phys. Rev. E \textbf{76}, 011113 (2007).
\bibitem{deGennes} P. G. de~Gennes and J. Prost, \textit{The Physics of Liquid Crystals} (Oxford University Press, 1993) [2nd edition, ISBN: 0198517858].
\bibitem{Review Casimir} M. Bordag, G. L. Klimchitskaya, U. Mohideen, and V. M. Mostepanenko, \textit{Advances in the Casimir Effect} (Oxford University Press, 2009) [ISBN-13: 978-0-19-923874-3].
\bibitem{warningjordan} Caution should be taken to avoid the case $\lambda_1=\lambda_2$, for which the dynamic matrix is not diagonalizable and a Jordan block appears. The method developed in this article is not directly applicable, but the generalization is simple. Also, neither $\lambda_1$ or $\lambda_2$ can vanish, because there is no damping term to make the nonconservative noise vanish and the fields would perform an unbounded random walk.
\bibitem{Elizalde} E. Elizalde, J. Phys. A Math. Gen. \textbf{27}, 3775 (1994).

\bibitem{LandauLifshitz} L. D. Landau and E. M. Lifshitz, \textit{Classical Theory of Fields} (Butterworth-Heinemann, 1980) [pp. 82-85, 4th edition, ISBN-13: 978-0750627689].
\bibitem{Buenzli-Soto} P. R. Buenzli and R. Soto, Phys. Rev. E. \textbf{78}, 020102(R) (2008).

\end{thebibliography}
\end{document}